# Compact orbital-angular-momentum multiplexing via laser-written glass chips

Chenhao Li[1,2], Simon Gross[3,4, *], Leonardo de S. Menezes[2, 5], Stefan A. Maier[6, 7, *], Judith M. Dawes[4, *], Haoran Ren[6, *]

[1]School of Electronic Science and Engineering, Xiamen University, Xiamen 361005, China

[2]Chair in Hybrid Nanosystems, Nanoinstitute Munich, Faculty of Physics, Ludwig Maximilian University of Munich, Munich 80539, Germany.

[3]School of Engineering, Macquarie University, NSW, 2109, Australia.

[4]MQ Photonics Research Centre, School of Mathematical and Physical Sciences, Macquarie University, NSW, 2109, Australia.

[5]Departamento de Física, Universidade Federal de Pernambuco, 50670-901 Recife-PE, Brazil.

[6]School of Physics and Astronomy, Faculty of Science, Monash University, Melbourne, Victoria 3800, Australia.

[7]Department of Physics, Imperial College London, London, SW7 2AZ, UK.

Correspondence emails:
simon.gross@mq.edu.au;
stefan.maier@monash.edu;
judith.dawes@mq.edu.au;
haoran.ren@monash.edu.

# Abstract

Orbital angular momentum (OAM) modes have emerged as a promising solution for enhancing the capacity of optical multiplexing systems, leveraging their theoretically unbounded set of orthogonal spatial modes. However, the generation and detection of OAM multiplexing signals are predominantly reliant on bulky optical components within complex optical setups. We introduce a compact solution for OAM information processing using laser-written glass chips, facilitating efficient multiplexing and demultiplexing of multiple OAM information channels. During the multiplexing process, OAM channels are managed via laser-scribed single-mode waveguides within a glass chip, with their modes converted using laser-written holograms on the side wall of the glass chip. The reverse optical process is employed for OAM demultiplexing. Our chips seamlessly interface with commercial optical fibers, ensuring compatibility with existing fiber-optic communication infrastructure. This work not only establishes a novel approach for OAM optical multiplexing but also underscores the potential of laser-writing technology in advancing photonics and its practical applications in optical communications.

**Introduction**

Optical multiplexing leverages various independent optical degrees of freedom, such as wavelength[1,2], polarization[3,4], and spatial modes[5], to enhance the capacity of both free-space and fiber-optic communication systems, thereby meeting the rapidly increasing demand for data transmission. Among these methods, space-division multiplexing has emerged as a vital approach for significantly increasing the bandwidth of multiplexing while maintaining generally low crosstalk[6,7]. The most promising approach within this framework involves the use of orbital angular momentum (OAM) modes, which are carried by a helical phase wavefront and offer a theoretically unbounded set of orthogonal modes[8]. Conventional OAM multiplexing typically relies on optical table systems to create and detect various OAM modes through bulky optical components, such as spatial light modulators[9,10], digital mirror devices[11], and spiral phase plates[12]. This reliance hinders system miniaturization. Although integrated photonic chips[13,14], ultrathin metasurfaces[15-17], and structured metafibers[18] have been developed to manipulate OAM beams, the operation of most previous approaches cannot be reversed and therefore focus only on either the generation or detection pathway.

Recent advancements in direct laser writing technology facilitate the precise scribing of optical waveguides within glass substrates [19,20]. By using a focused ultrafast laser beam, this technique allows for highly localized modifications of the refractive index, creating a low-loss pathway for light to travel through the glass. Meanwhile, 3D laser nanoprinting via two-photon polymerization (TPL) empowers the fabrication of arbitrary 3D nanostructures with feature sizes below the optical diffraction limit [21,22]. TPL facilitates the creation of functional photonic structures, such as refractive[23], diffractive[24] and metasurface[25] optical elements, as well as hybridized devices[26], on a diverse range of substrates such as glass[27], optical fibers[28], photonic chips[29], and imaging sensors[30]. As a result, direct laser writing serves as a powerful tool for developing miniaturized optical systems.

Here, we demonstrate compact OAM (de)multiplexing using laser-written glass chips. These chips feature OAM (de)multiplexing holograms on their end-faces, as well as laser-scribed single-mode waveguides within the volume of the glass chips. Operating at the standard communication wavelength of 1550 nm, our devices facilitate reversible information transmission. They allow spatial (de)multiplexing of four distinct OAM modes, specifically topological charges of ±2 and ±4, achieving efficient OAM information processing without the need for additional optical components (Fig. 1a). OAM (de)multiplexing phase holograms were 3D laser printed onto the glass chips' end-faces with photosensitive polymers. Four optical waveguides were laser-scribed within each glass chip to generate and detect the OAM multiplexing signals. The positions and angles of these waveguides were optimized to couple light to and from the OAM multiplexing holograms. These waveguides were then interfaced with standard telecommunication optical fibers, offering efficient light coupling and high immunity to spatial misalignment. Our OAM (de)multiplexing glass chips were pigtailed to standard optical fiber arrays and packaged in a custom aluminum enclosure for robustness (Fig. 1b). This demonstration presents a new approach to OAM multiplexing technology and highlights the potential of direct laser writing for photonic integration and information processing.

**Results**

To offer all-on-glass generation and detection of the OAM modes, four single-mode optical waveguides were first inscribed by an ultrafast laser (Femtolasers Femtosource 500XL) using the femtosecond laser direct-write technique[19, 31], as illustrated in Fig. 2a. An incident laser beam at a wavelength of 800 nm with <50 fs pulse during was focused 0.55 mm below the top surface of a 1.1 mm thick boro-aluminosilicate glass sample (Corning EagleXG) using a 20×, 0.45 numerical aperture microscope objective (Olympus LUCPLAN-N). The repetition rate of the laser was set to 509 kHz using an external electro-optical pulse picker and the pulse energy reduced to 100 nJ as measured in front of the microscope objective. The multiscan technique[32] was employed to create approximately square-shaped waveguides with a cross-section of 8 by 8 µm$^2$ via the consecutive side-by-side inscription of 16 individual modifications at a pitch of 0.5 µm while translating the sample at a velocity of 1 mm/s. The inscription order of the 16 modifications was according to the half-scan method[33] with a half-scan pitch of 1.5 µm. The refractive index contrast of the waveguides was measured to be 3.2×10$^{-3}$ by inscribing 20 µm short waveguide stubs using identical parameters[34]. The refractive index contrast characterisation was conducted at a wavelength of 600 nm using a commercial camera (PHASICS SID4BIO) that employs quadri-wave lateral shearing interferometry to spatially reveal changes in the optical thickness of transparent materials[35]. All waveguides were designed for single-mode operation at 1550 nm with a measured mode-field diameter of 13.0 × 13.2 µm (horizontal × vertical).

To interface the glass chips with optical fibers, the four waveguides were arranged as a linear array spaced by 127 µm and coupled to standard single-mode optical fiber arrays. For illumination of the hologram on the end-face of the glass chip, the four waveguides were spatially arranged at the vertices of a square pattern with a side-length of 105 µm (inset of Fig. 2a). Each waveguide was approaching their termination position under a tilt of 0.75 degrees, in order for every waveguide output to illuminate exactly the same spot on the end-face of the glass chip, where an OAM multiplexing hologram was later fabricated. The waveguides terminated 7.7 mm away from the end-face of the glass chip, ensuring that the waveguide outputs sufficiently diverge to cover the entire area of the OAM multiplexing hologram. Notably, the divergent and tilted wavefronts from the four waveguide outputs can be compensated by a lens and a grating function added to the OAM (de)multiplexing hologram, respectively (Supplementary Fig. 1). This configuration allows for the co-axial generation of four OAM beams. Moreover, this optimized design enhances the coupling of OAM-encoded signals into the waveguides during the demultiplexing stage, thereby improving overall transmission efficiency. Figure 2b presents the mode distributions of the four individual waveguides, which were experimentally imaged from the end-face of one of the glass chips. The results indicate that the waveguide outputs achieve nearly perfect spatial overlap, with each channel exhibiting a Gaussian-shaped fundamental mode. We further measured the transmission of the fiber-coupled waveguides, as shown in Fig. 2c, revealing a low loss of about -1.7 dB, which includes waveguide propagation losses, coupling losses between waveguides and fibers, connector losses and 0.18 dB Fresnel reflection at the glass chip's end-face. The insertion losses exhibit excellent consistency across all four channels. This transmission consistency ensures that no additional corrections were needed for the on-glass processing of the OAM-encoded signals.

The second key component of the glass chip is a 3D laser-printed phase hologram designed for implementing OAM (de)multiplexing (Fig. 3a). Four spiral phase plates with OAM modes of ±2 and ±4 were equipped with distinctive linear gratings. In the Fourier plane of the hologram, these gratings induce spatial shifts of the OAM modes in both x and y directions, precisely aligning with the ends of the four waveguides inside the glass (Supplementary Fig. 2). During the OAM multiplexing process, the angular shifts introduced by the gratings are compensated by the tilted incidence from the four waveguide outputs. This ensures that all four OAM modes are generated co-axially after passing through the OAM multiplexing hologram (Supplementary Figs. 1c and 1d). A Fourier lens profile with a focal length into glass of 7.7 mm was added to the OAM multiplexing hologram. This addition allows the waveguide outputs to be transformed into collimated beams (beam diameter of 1 mm) during the multiplexing process. Meanwhile, it ensures that OAM incident beams are efficiently coupled into the waveguides during the demultiplexing process, due to their matched numerical aperture with the single-mode waveguides. Thus, the four OAM modes, at normal incidence on the OAM multiplexing hologram, can be selectively converted back to the fundamental Gaussian mode and coupled to the individual single-mode waveguides (Supplementary Figs. 1e and 1f).

Our OAM (de)multiplexing hologram was designed with 2000 by 2000 pixels, each with a pixel pitch of 500 nm, resulting in an aperture diameter of 1 mm. This hologram was fabricated through a commercial direct laser writing system (Nanoscribe GT2) in a photoresist material (IP-L) with a refractive index close to 1.5 at a wavelength of 1550 nm (see Methods). Given the refractive index difference $\Delta n = 0.5$ between the polymer and air at the operating wavelength of 1550 nm, the maximum height difference of the hologram pixels was designed to be 3.1 μm to cover the full $2\pi$ phase modulation. We selected 16 different heights to evenly cover the 0 to $2\pi$ phase modulation. Figure 3b presents optical and SEM images of the 3D laser-printed hologram. We used a home-built optical imaging setup (Supplementary Fig. 3) to characterize the generated OAM modes by the multiplexing hologram. The OAM modes were collected in the Fourier plane of the hologram through individually illuminating the four input fibers. The imaging results were recorded by a short-wave infrared camera (Owl 640 M), as shown in Supplementary Fig. 4. Figure 3c presents the intensity distributions of four OAM output beams collected at the end-face of a glass chip. To further confirm the specific OAM modes, we adopted the astigmatic transformation method using a tilted spherical lens[36]. The number of fringes indicates the OAM order, while the deflection orientation differentiates the positive and negative orders. Notably, the designed four OAM modes (±2 and ±4) were successfully generated from the OAM multiplexing glass chip, with the modes from the four waveguides achieving good spatial overlap at the end facet of the glass chip.

The OAM multiplexing glass chip can also be used for the demultiplexing process through the reversible operation of the glass chip. In this process, when incident light carrying a specific

OAM mode reaches the hologram, it splits into four directions toward the four waveguides. Due to OAM conservation, only one output beam from the hologram can be selectively converted to the fundamental Gaussian mode, which is further coupled into the specific waveguide. Other non-zero OAM modes with doughnut-shaped intensities at the remaining waveguides are rejected by the single-mode waveguides due to coupling mismatches. The simulated demultiplexing results are given in Supplementary Fig. 5. We firstly used the OAM multiplexing hologram fabricated on a planar silica substrate to verify the OAM demultiplexing principle (Supplementary Fig. 6). When illuminated by four different OAM incident beams at normal incidence, the imaging results at the Fourier plane (focal distance of 7.7 mm) of the hologram showed great consistency with the simulated demultiplexing results. Following these initial validations, all-on-glass OAM multiplexing and demultiplexing experiments were performed by the combined use of two fabricated glass chips (Fig. 4a). We used an optical imaging system with two telescopes to precisely align the end-faces of two glass chips (blue box in Fig. 4a), which can be removed once the alignment work is completed. On the Fourier plane of the first telescope, we placed a pinhole to block the off-axis OAM modes. As a result, different OAM modes ($\pm 2$ and $\pm 4$) generated from the OAM multiplexing glass chip were incident on the OAM demultiplexing glass chip. These OAM demultiplexing signals were detected by the four fiber outputs, showing good intensity contrasts for the desired OAM modes (Fig. 4b). Our demonstrated OAM multiplexing and demultiplexing glass chips can be readily used for practical OAM-based optical communications. Figure 4e displays the crosstalk matrix of the OAM (de)multiplexing results based on our glass chip protocol. It shows that the transmitter successfully encoded optical signals onto the selected OAM modes, which were accurately decoded by the receiver, achieving an average crosstalk of -9.6 dB. Although our simulated crosstalk can be as low as -20 dB, the increased crosstalk can be attributed to fabrication imperfections in both the holograms and optical waveguides, as well as some alignment errors.

**Conclusion**

We have demonstrated fully integrated glass chips for compact OAM multiplexing and demultiplexing, offering a practical solution for free-space optical communication links. Our all-glass chips efficiently multiplex and demultiplex four OAM modes, achieving an average modal crosstalk of -9.6 dB. Using ultrafast laser processing, we inscribed single-mode waveguides within the glass substrate, enabling precise waveguide positioning, controlled waveguide outcoupling directionality, and seamless integration with optical fibers. Furthermore, we employed 3D direct laser writing technology to implement the OAM (de)multiplexing phase holograms directly on the end-faces of the glass substrates. This facilitates high-precision generation (multiplexing) and detection (demultiplexing) of OAM modes on a chip. The dual application of laser technologies results in robust, streamlined photonic packaging. The chip's high integration level significantly reduces its size, making it an ideal solution for compact optical communication systems. In future work, we aim to increase the number of OAM modes for optical multiplexing and to combine OAM multiplexing with other degrees of freedom of light, such as polarization and wavelength. We believe our demonstration offers a viable approach for all-on-chip OAM (de)multiplexing, paving the way for next-generation optical communications and information processing.

**Methods**

**3D laser nanoprinting of OAM multiplexing hologram on planar silica substrates**

The OAM multiplexing hologram was fabricated on a silica substrate using IP-L 780 photoresist resin (Nanoscribe GmbH) through a Nanoscribe GT two-photon polymerization lithography system. A high numerical aperture (NA) objective (Plan-Apochromat 63×/1.40 Oil DIC, Zeiss) was utilized in an immersion configuration to achieve precise structural definition. The fabrication parameters were optimized to a laser power of 45 mW and a scanning speed of 10,000 μm/s. The total writing time for an OAM multiplexing hologram was 1.2 hour. After laser exposure, the samples were sequentially developed in propylene glycol monomethyl ether acetate (PGMEA, Sigma-Aldrich) for 20 minutes, isopropyl alcohol (IPA, Sigma-Aldrich) for 5 minutes, and methoxy-nonafluorobutane (Novec 7100 Engineered Fluid, 3M) for 2 minutes. The final samples were dried via air evaporation. The scanning mode was set to galvo mirror scanning to increase the fabrication speed, the hologram was segmented into square cells measuring 100 μm × 100 μm.

**3D laser nanoprinting of OAM multiplexing holograms on glass chips**

The waveguide glass is encapsulated within a custom-designed metal housing, enabling the entire sample to be mounted into the Nanoscribe GT system. Precise alignment is required to ensure the hologram's printing position corresponds accurately to the waveguide position. To facilitate this alignment, four reference waveguides were incorporated during the laser-engraving process on the waveguide glass. These reference waveguides were arranged in a cross formation, with a spacing of 1 mm between adjacent waveguides, and their outlet positioned 50 μm from the surface. During hologram fabrication, the reference waveguides were illuminated using a fiber optic illuminator. Their positions were then identified by the camera of the two-photon lithography system, allowing the printing position to be precisely set at the intersection of the cross. The processing flow and fabrication recipe for the waveguide glass were kept identical to those used for planar silica substrates.

**Acknowledgements**

J. D., S. A. M., and H. R. acknowledge funding support from the Australian Research Council (DP220102152). H. R. acknowledges the DECRA (DE220101085) funding support from the Australian Research Council. S. G. acknowledges funding by an ARC Future Fellowship (FT200100590). S. A. M. acknowledges funding support from the Lee Lucas Chair in Physics. C. Li and L. de S. Menezes acknowledge the support of the Center for Nanoscience (CeNS), Ludwig-Maximilians-Universität München. This work was performed in part at the Melbourne Centre for Nanofabrication (MCN) in the Victorian Node of the Australian National Fabrication Facility (ANFF).

**Contributions**

J. D., S. A. M., and H. R. conceived the idea; C.L. performed the numerical analysis, fabrication, and experimental characterization; L. de S. M. supported the imaging setup. S. G. contributed to the waveguide glass chips; C. L., S. G., L. M., S. A. M., J. D., and H. R. contributed to data analysis; C. L. and H. R. wrote the paper draft with contributions from all authors.

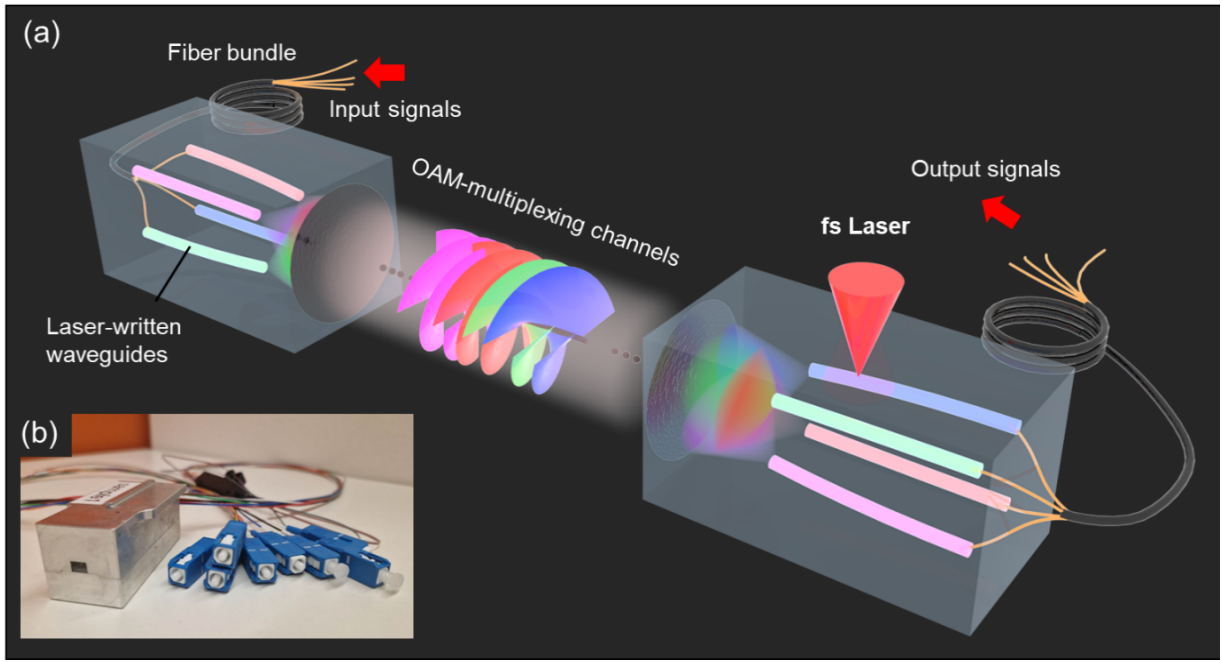

**Figure 1. Compact OAM multiplexing and demultiplexing via laser-processed glass chips.** (**a**) Schematic illustration of the OAM multiplexing and demultiplexing using a reversible spatial arrangement of two glass chips interfaced with fiber bundles. (**b**) Photograph of a fabricated glass chip encased in a metal housing for protection.

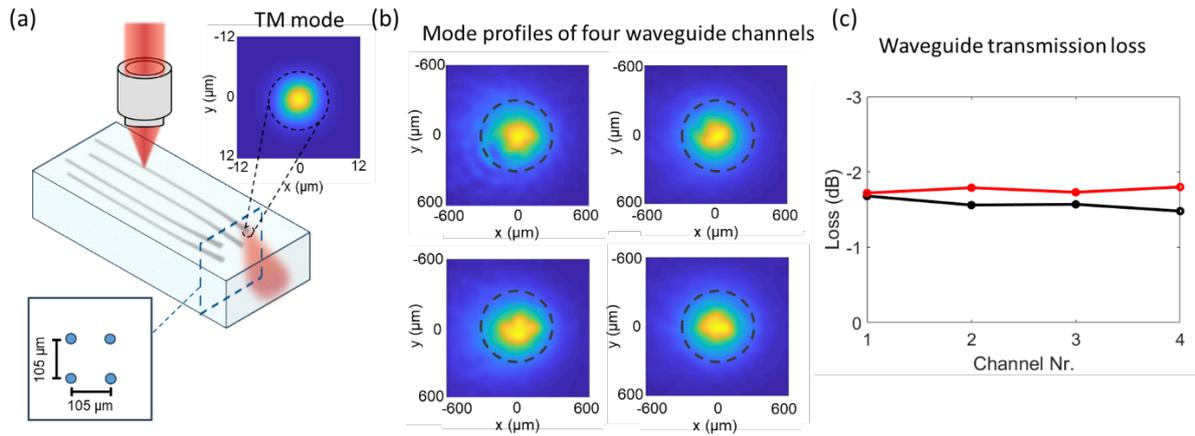

**Figure 2. Optical performance of four laser-inscribed single-mode waveguides.** (**a**) Schematic diagram of laser-inscribed glass waveguides. The grey lines denote the four inscribed single-mode waveguides, with their supported transverse magnetic (TM) mode profile in the inset. The dashed blue rectangle indicates the spatial arrangement of the four waveguide outputs. (**b**) Mode profiles of the four fabricated single-mode waveguides. The dashed black circles denote the area for the OAM (de)multiplexing hologram in a later stage. (**c**) Transmission loss analysis of the four waveguides.

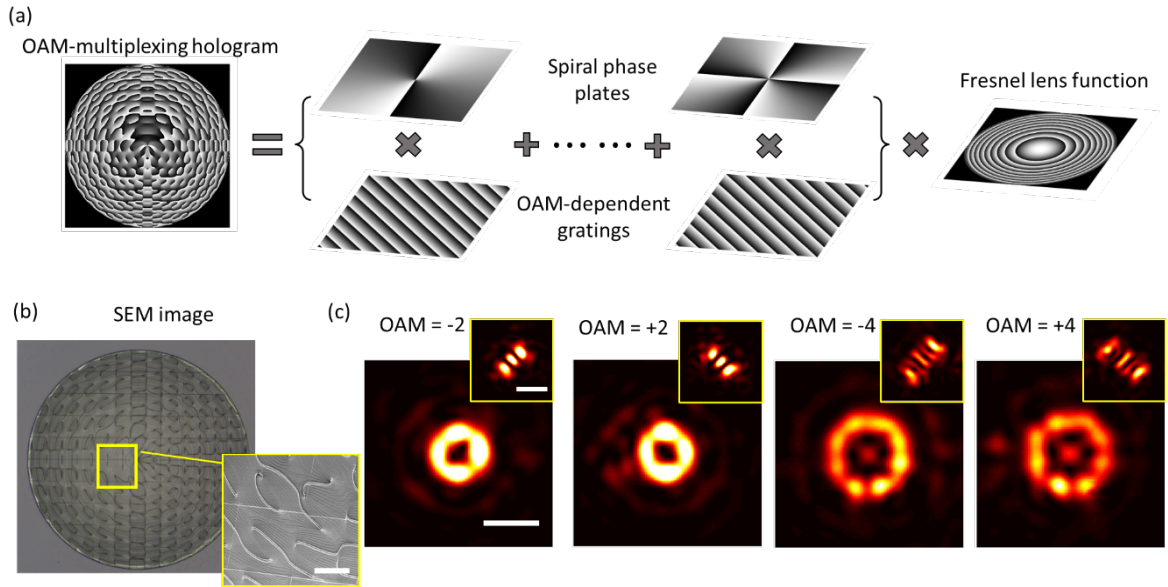

**Figure 3. Design and experimental characterization of an OAM-multiplexing hologram.** (**a**) Illustration of the OAM-multiplexing hologram design. (**b**) Optical and SEM (inset) images of the fabricated hologram device (scale bar: 200 nm). (**c**) Experimental characterization of the hologram (scale bar: 2 µm). Top right: The astigmatic transformation patterns induced by a tilted spherical lens. The number of dark fringes indicates the OAM order, while their orientation indicates a positive or negative OAM order.

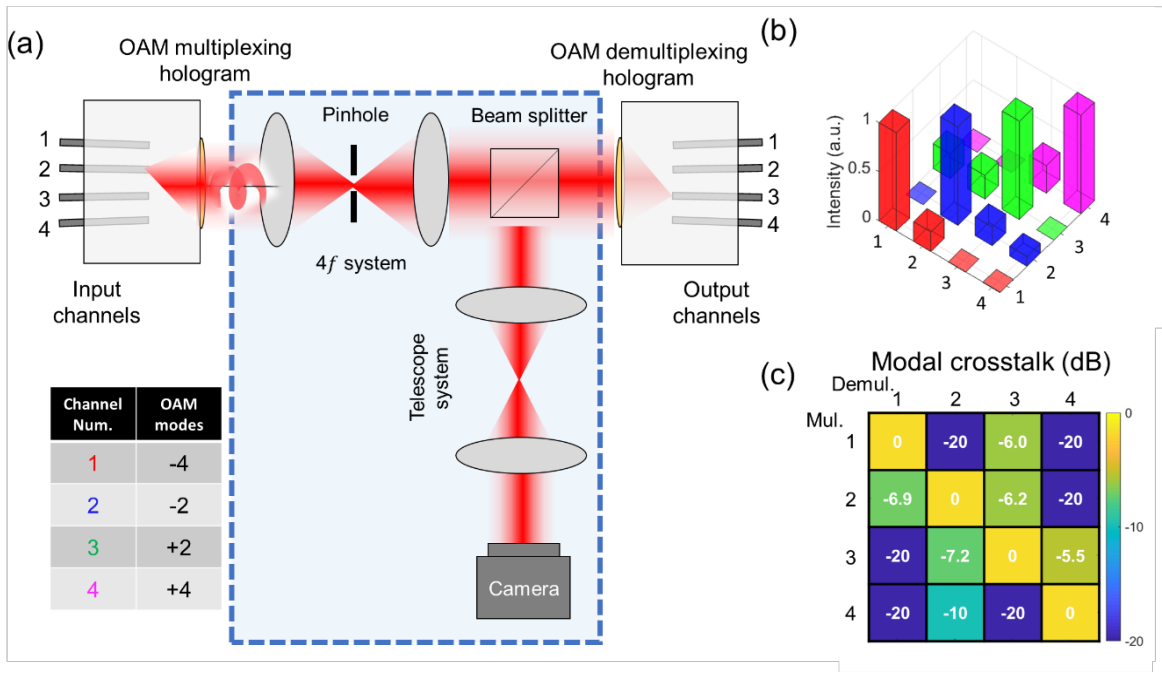

**Figure 4. All-on-glass OAM multiplexing and demultiplexing.** (**a**) Optical setup used for aligning the OAM multiplexing glass chip with the OAM demultiplexing glass chip. The blue box indicates the alignment imaging setup, which can be removed after alignment. (**b**) Experimental characterization of the OAM output signals from the OAM demultiplexing glass chip based on the OAM input signals from the OAM multiplexing glass chip. (**c**) Multiplexing crosstalk presented as the OAM mode matrix.